\documentclass[prx,twocolumn,showpacs,preprintnumbers,amsmath,amssymb]{revtex4-1}

\usepackage{graphicx,natbib}
\usepackage{dcolumn}
\usepackage{bm}
\usepackage{color}

\begin{document}

\title{Vortex Mass in a Superfluid}
\author{Tapio Simula}
\affiliation{School of Physics and Astronomy, Monash University, Victoria 3800, Australia}

\begin{abstract}
{We consider the inertial mass of a vortex in a superfluid. We obtain a vortex mass that is well defined and is determined microscopically and self-consistently by the elementary excitation energy of the kelvon quasiparticle localised within the vortex core. The obtained result for the vortex mass is found to be consistent with experimental observations on superfluid quantum gases and vortex rings in water. We propose a method to measure the inertial rest mass and Berry phase of a vortex in superfluid Bose and Fermi gases.}
\end{abstract}

\maketitle

\section{Introduction}
The mass of a vortex in a superfluid has been debated in the literature for some time with predictions ranging from it being practically zero or infinite to being not well defined. The vortex mass candidate, $M_{\rm field}=E_{\rm field}/c_s^2$, where $c_s$ is the speed of sound, as noted by Popov \cite{Popov1973a}, Duan and Leggett \cite{Duan1992a,Duan1994a} and many others, seems to result in a logarithmically divergent vortex mass for large distances $R$ relative to the vortex core size $r_c$ because the incompressible kinetic energy $E_{\rm field}$ associated with the superflow velocity field of a static vortex is proportional to $\ln({R/r_c})$. An alternative approach due to Baym and Chandler \cite{Baym1983a} is to consider not the energy to create the vortex itself but the kinetic energy cost to move it. This seems to result in a negligible vortex mass equal to the bare mass $M_{\rm bare}=\rho_s \pi r_c^2L$, where $\rho_s$ is the superfluid mass density and $L$ the length of the vortex, that essentially corresponds to the mass of the fluid displaced by the vortex core \cite{Baym1986a}. In Fermi superfluids an induced hydrodynamic vortex mass, known as Kopnin mass, has been found to be associated with quasiparticle bound states trapped within the vortex core \cite{Kopnin1978a,Sonin1987a,Volovik2003a,Sonin2013a}. 

Thouless and Anglin considered a vortex pinned by an external potential that moves in a circular path and concluded that the concept of the vortex mass would not be well defined and would depend on the details of the measurement process \cite{Thouless2007a}. However, application of such a pinning potential strongly affects the Kelvin modes localised at the vortex core \cite{Pitaevskii1961a,Gross1961a,Isoshima1999a,Simula2002a,Simula2008a,Simula2010a}, which means that in the presence of a pinning potential the vortex mass cannot be decoupled from the properties of the applied pinning potential even if the vortex would be dragged through the superfluid adiabatically \cite{Simula2001a}. We are able to avoid this subtle issue by allowing the vortex to move free from localised external potentials, which makes it possible to relate the inertial vortex mass to the frequency of the elementary kelvon excitation. Furthermore, if one considers self-trapped cold atom gases held together by long-range particle interactions one may avoid using any external potentials altogether \cite{Schmitt2016a}.

Here we approach this conundrum by considering a self-propelling vortex that moves in a superfluid Bose--Einstein condensate spontaneously along a circular orbit due to the self-induced asymmetry in the superflow. Observing the vortex motion from the rotating frame of reference leads to an unambiguous definition of the inertial mass of such a vortex in terms of the energy of the unique kelvon quasiparticle of the elementary excitation spectrum of the superfluid. The kelvon is the finger print of the quantised vortex and its excitation energy is determined by a self-consistent theory that contains the information pertinent to the microscopic origin of the vortex mass in a Bose--Einstein condensate.

We obtain an all-inclusive (total) inertial mass of a vortex 
\begin{equation}
M_{\rm vortex} = \frac{\Gamma \rho}{\omega_{\rm K}} L,
\label{classicmass}
\end{equation}
where $\Gamma$ is the circulation of the vortex, $\rho$ is the mass density of the fluid hosting the vortex, $\omega_{\rm K}$ is the angular frequency of the fundamental Kelvin wave excitation of the vortex, and $L$ is its length. We conclude that a vortex in a superfluid is heavy and its mass is finite and well defined spectroscopically by two measurables; the condensate density and the frequency of the kelvon quasiparticle. We suggest how the Berry phase, the Magnus force, the vortex velocity dipole moment (vVDM), and the inertial rest mass of a vortex could be measured experimentally in superfluid Bose and Fermi gases.

In what follows, on discussing various mass candidates we use upper case symbols such as $M$ to denote a three-dimensional mass and lower case symbols such as $m=M/L$ to denote two-dimensional mass or mass per unit length. The particle densities and mass densities appearing in the vortex mass definitions are three-dimensional. Strictly, the fluid density appearing in our result for the inertial mass of a vortex in a superfluid derived for dilute gas Bose--Einstein condensates at low temperatures is the condensate particle density. However, in strongly interacting superfluids such as helium II, the condensate density is not easily experimentally observable and in such cases, we use the condensate mass density and the superfluid density interchangeably although in helium II the latter may exceed the condensate density by an order of magnitude. 

In Sec.~II we outline the theoretical description of weakly interacting superfluids in terms of Bogoliubov quasiparticles noting the qualitative similarities between Bose and Fermi systems at the quasiparticle level. In Sec.~III we provide a derivation of the equation of motion of a quantised vortex in a Bose--Einstein condensate. The result shows that the only term contributing to the vortex velocity is the Laplacian in the Gross--Pitaevskii equation and that there is no explicit transverse force term acting on the vortex due to the motion of the thermal cloud. Sections ~IV and V are devoted for discussing, respectively, the properties of Kelvin waves of vortices and their quantum mechanical counterparts---the kelvons. In Sec.~VI we derive the equation for the inertial mass of a vortex and show that it is determined by the excitation frequency of the fundamental kelvon of the vortex. In Sec.~VII we elucidate the origin of the vortex mass in terms of the kelvon quasiparticle and compare the quasiparticle excitation spectra due to the structure of the vortex core in Bose and Fermi gases. The connection between the geometric Berry phase and the inertial mass of a vortex is also provided.  In Sec.~VIII we discuss how the vortex mass and its Berry phase could potentially be measured in cold atom experiments. In Sec.~IX the classical limit of large quantum numbers is considered showing how the formula for the vortex mass may be applicable to classical fluids such as water. Section X concludes this work.

\section{Quasiparticle picture}
Our starting point for the derivation of the vortex mass is the dynamical Hartree--Fock--Bogoliubov theory of the superfluid Bose gas, which is known to yield an elementary excitation spectrum that agrees with the second order accurate Beliaev theory \cite{Griffin2009a}. In this theory the dynamics of the condensate wave function $\psi({\bf r}, t)$ are determined by the generalised Gross--Pitaevskii equation
\begin{equation}
i\hbar\partial_t \psi({\bf r}, t) = \mathcal{L}({\bf r}, t)   \psi({\bf r}, t)  +g\Delta({\bf r}, t)  \psi^*({\bf r}, t)
\label{GP}
\end{equation}
where
$
\mathcal{L}({\bf r}, t) =- \frac{\hbar^2}{2M}\nabla^2   +V_{\rm ext}({\bf r},t)  +  gn({\bf r}, t)+2g\rho({\bf r}, t) -iD({\bf r},t).
$
In Eq.~(\ref{GP}), 
$M$ is the mass of a particle the superfluid is composed of, $g$ determines the strength of the $s$-wave interactions in the system, $n({\bf r}, t)=|\psi({\bf r}, t)|^2$, $\rho({\bf r}, t)$ and $\Delta({\bf r}, t)$ are the condensate, non-condensate and pair-potential densities, respectively, $V_{\rm ext}({\bf r},t) $ includes all external pinning, trapping and other potentials and the non-Hermitian term $iD({\bf r},t)$ accounts for condensate growth and loss processes such as dissipation due to the interaction with non-superfluid atoms.

The internal potentials $\rho({\bf r}, t)$ and $\Delta({\bf r}, t)$ involve summations over all quasiparticle states and are determined \emph{self-consistently} by the relations
\begin{align}
\rho({\bf r}, t) = \sum_{qp} \big\{ f_{qp}(t) [u^*_q({\bf r}, t)u_p({\bf r}, t) + v^*_q({\bf r}, t)v_p({\bf r}, t) ] \notag \\
                     -2{\rm Re} [g_{qp}(t) u_q({\bf r}, t)v_p({\bf r}, t)] + \delta_{qp}|v_q({\bf r}, t)|^2 \big\}
                     \label{rho}
\end{align}
and
\begin{align}
\label{delta}
\Delta({\bf r}, t) = -&\sum_{qp} \big\{ [2f_{qp}(t)+\delta_{qp}] v^*_q({\bf r}, t)u_p({\bf r}, t)  \\
                     -&g_{qp}(t) u_q({\bf r}, t)u_p({\bf r}, t) -g^*_{qp}(t) v^*_q({\bf r}, t)v^*_p({\bf r}, t)\big\}. \notag
\end{align}
The last term in $\rho({\bf r}, t)$ is proportional to the Lee--Huang--Yang contribution to the ground state energy due to the quantum fluctuations \cite{Lee1957a} and it yields a non-zero quantum depletion even at zero temperature, the effects of which have recently been observed experimentally \cite{Schmitt2016a,Chang2016a}. The evolution of the quasiparticle amplitudes $u_q({\bf r}, t)$ and $v_q({\bf r}, t)$ are determined by the Bogoliubov--de Gennes (BdG) equations
\begin{align}
i\hbar\partial_t u_q({\bf r}, t) &=&\mathcal{M}({\bf r}, t) u_q({\bf r}, t)       -  \mathcal{V}({\bf r}, t)   v_q({\bf r}, t)\notag \\
i\hbar\partial_t v_q({\bf r}, t) &=& -\mathcal{M}^*({\bf r}, t) v_q({\bf r}, t)  + \mathcal{V}^*({\bf r}, t) u_q({\bf r}, t)
\label{bog}
\end{align}
expressed in terms of the operators
\begin{align}
\mathcal{M}({\bf r}, t) =   - \frac{\hbar^2}{2M}\nabla^2 -\mu   +V_{\rm ext}({\bf r},t)  +2gn({\bf r}, t) +2g\rho({\bf r}, t)  \notag
\label{Mpot}
\end{align}
and
\begin{align}
\mathcal{V}({\bf r}, t) = g\psi^2({\bf r}, t) + g\Delta({\bf r}, t),\notag 
\end{align}
where $\mu$ is the chemical potential. The quasiparticle distribution functions $f_{qp}(t)=\langle  \alpha^\dagger_q \alpha_p\rangle$ and $g_{qp}(t)=\langle  \alpha_q \alpha_p\rangle$ are expectation values of products of quasiparticle creation $ \alpha^\dagger_q$ and annihilation $\alpha_q$ operators.

Fermi superfluids may be modelled using similar quasiparticle picture with two major differences. A coherent condensate mode $\psi({\bf r})$ described by the Gross--Pitaevskii equation, Eq.~(\ref{GP}), is absent and the quantum statistics of the quasiparticles of Eqns.~(\ref{bog}) are characterised by fermion anticommutation relations instead of boson commutation relations. The gap function $\Delta({\bf r})$ replaces the condensate mode as the relevant order parameter in this case. Importantly, however, for both Bose and Fermi systems the BdG equations have topologically non-trivial vortex solutions characterised by emergent quasiparticle states at the vortex core. For Bose and Fermi systems these vortex core localised quasiparticle states are the kelvons \cite{Pitaevskii1961a} and Caroli--Matricon--deGennes (CdGM) states \cite{Caroli1964a,Virtanen1999a,Mizushima2008a,Prem2017a}, respectively.

\section{Vortex velocity and the transverse force}

To obtain the equation of motion for an isolated quantised vortex in a two-dimensional Bose--Einstein condensate, we may consider a generic vortex state 
\begin{align}
\psi_v({\bf r}, t) =f({\bf r})  \sqrt{e({\bf r}, t)}e^{iS({\bf r},t)}
\end{align}
, where $S({\bf r},t)$ is a smooth real function, $e({\bf r}, t)$ is defined by $|\psi_v({\bf r}, t)|^2=|{\bf r}-{\bf r}_v|^2e({\bf r}, t) $, and the function $f({\bf r})$ accounts for the non-analytic internal structure within the vortex core. The exact form of $\psi_v({\bf r}, t)$ is determined self-consistently by Eq.~(\ref{GP}). In general, the structure function $f({\bf r})$ contains multipole moments to all orders but for the purpose of deriving the vortex velocity equation it is sufficient to approximate it by only considering the lowest order monopole field
\begin{align}
f({\bf r})= x-x_v + i(y-y_v),
\end{align}
where $(x_v,y_v)$ are the position coordinates of the vortex. The influence of the dipolar velocity field due to the kelvon quasiparticle is limited to within the vortex core and yields the vortex velocity dipole moment (vVDM) \cite{Klein2014a,Groszek2017a}.
The equation of motion of the vortex is obtained explicitly by linearising the formal solution of the generalized time-dependent Gross-Pitaevskii equation, Eq.~(\ref{GP}), and finding the position ${\bf r}_v(t+\delta t)$ of the phase singularity of the vortex wavefunction $\psi_v({\bf r}, t+\delta t)$, where $\delta t$ is an infinitesimal time increment \cite{Groszek2017a}. The resulting velocity of the vortex ${\boldsymbol v}_v({\bf r}_v, t)\equiv\delta{\bf r}_v/\delta t$ is 
\begin{equation}
{\boldsymbol v}_v({\bf r}_v, t) = {\boldsymbol v}_s({\bf r}_v, t) - \frac{\hbar\hat{\bf e}_z\times\nabla e({\bf r}, t) }{2Me({\bf r}, t)}\bigg|_{{\bf r}_v},
\label{vortexvelo}
\end{equation} 
where ${\boldsymbol v}_s({\bf r}_v, t)=\kappa \nabla S({\bf r},t)/2\pi|_{{\bf r}_v}$ is the embedding velocity in the vicinity of the vortex core and $\kappa=h/M$ is the quantum of circulation. The vortex velocity equation (\ref{vortexvelo}) depends on the gradients of the embedding phase and the condensate density in the vicinity of the vortex and provides a complete description of the vortex motion including effects such as vortex `friction' that results, for example, in radial drifting of the vortex in the presence of noncondensate atoms in harmonic traps \cite{Rooney2010a}. 

Equation (\ref{vortexvelo}) may be multiplied from left by $[-M\tilde{n}({\bf r}_v, t) {\boldsymbol \kappa}\times]$, where ${\boldsymbol \kappa}=\kappa\hat{\bf e}_z$ and $\tilde{n}({\bf r}_v, t)$ is the condensate density in the absence of the vortex. This yields a force balance equation
\begin{equation}
{\bf f}_{v} =  {\bf f}_{s}  + {\bf f}_{\rm buoy}, 
 \label{forcebalance}
\end{equation} 
where the lower case symbols $\bf{f}$ denote force per unit length such that the total force ${\bf F}_{v}$ on a columnar vortex of length $L$ is
\begin{equation}
{\bf F}_{v} = {\bf f}_{v} L.
\end{equation} 
Equation ~(\ref{forcebalance}) may also be expressed as
\begin{equation}
{\bf f}_{\rm Mag} + {\bf f}_{\rm buoy} =0
 \label{forcebalancemagnus}
\end{equation} 
in terms of the Magnus force, also known as the Kutta--Joukowski lift
\begin{equation}
{\bf f}_{\rm Mag}=M\tilde{n}({\bf r}_v, t) {\boldsymbol \kappa}\times({\boldsymbol v}_v- {\boldsymbol v}_s)={\bf f}_{s}-{\bf f}_{v}.
\label{Magnus}
\end{equation} 
The gradient force due to the inhomogeneous condensate density, 
\begin{equation}
{\bf f}_{\rm buoy} = -\frac{\pi\hbar^2}{M} \frac{\tilde{n}({\bf r}, t)}{e({\bf r}, t)}\nabla e({\bf r}, t) \bigg|_{{\bf r}_v},
\label{fbuoy}
\end{equation} 
originates from the quantum pressure and can in general point in any direction causing the vortex to accelerate or decelerate. 

In Eq.~(\ref{forcebalance}) there is no \emph{explicit} transverse force on the vortex due to the motion of the non-condensate. Instead, all such effects involving the non-condensate  are included in the vortex dynamics implicitly and self-consistently through ${\bf f}_{\rm buoy}$ and ${\bf f}_{\rm s}$  via their influence on the density $n({\bf r}, t)$ and embedding velocity ${\boldsymbol v}_s({\bf r}, t)$ of the \emph{condensate} wave function, $\psi({\bf r}, t)$ in Eq.~(\ref{GP}). 

\section{Kelvin waves}
Lord Kelvin studied small amplitude perturbations to a thin columnar vortex with hollow core and obtained a dispersion relation
\begin{equation}
\omega_{\rm K} = \frac{\Gamma}{2\pi r_c^2} \left( 1 - \sqrt{1+kr_c\left[ \frac{K_0(kr_c)}{K_1(kr_c)}\right]}    \right),
\label{Kelvinbessel}
\end{equation} 
where $\Gamma$ is the circulation, $r_c$ is the core parameter of the vortex, $k$ is the wave vector and $K_j$ is a modified Bessel functions of order $j$ \cite{Kelvin1880a}. These Kelvin waves correspond to infinitesimal perturbations to the inner surface of the vortex core and manifest as propagating helical displacement of the centre line of the vortex core. Remarkably, such Kelvin waves may be amplified to the extent that the amplitude of the perturbation becomes greater than the core size of the vortex allowing clear visualisation of the deformed helical shape of the vortex. The Kelvin wave dispersion relation Eq.~(\ref{Kelvinbessel}) has a long wave length approximation
\begin{equation}
\omega^L_{\rm K} = \frac{\Gamma k^2}{4\pi}\left[\log\left( \frac{2}{kr_c}\right) -\gamma\right],
\label{Kelvinlog}
\end{equation} 
where $\gamma$ is the Euler--Mascheroni constant.

 Pocklington studied sinuous waves on a hollow vortex ring \cite{Pocklington1895a} obtaining the Kelvin wave dispersion relation, Eq.~(\ref{Kelvinbessel}). However, its validity is then limited to great wavenumbers such that the shape of the vortex ring appears rectilinear on length scales comparable to the wavelength of the perturbation. Thomson considered instead the slowly varying Kelvin waves on a vortex ring finding a dispersion relation \cite{Thomson1883a} 
\begin{equation}
\omega^R_{\rm K} = \frac{\eta V_R}{R},
\label{KelvinThom}
\end{equation} 
expressed in terms of the radius $R$ of the ring and its translational speed 
\begin{equation}
V_R =   \frac{\Gamma}{ 4\pi R} \left[\log\left(\frac{8R}{r_c}\right) - \beta\right], 
\label{ringspeed}
\end{equation}
where the value of $\beta$ depends on the details of the structure of vortex core. The factor $\eta = \sqrt{p^2(p^2-1)}\approx p^2 -\frac{1}{2}$ is expressed in terms of the integer $p=1,2,3\ldots$ and the latter form is the best quadratic polynomial approximation. The two forms differ significantly only for the longest wave length mode $p=1$. 

\section{Kelvons}
Quantised vortices in superfluids are in many respects similar to their classical counterparts. Pitaevskii studied the small amplitude perturbations to quantised vortex lines \cite{Pitaevskii1961a} and obtained a dispersion relation equivalent to Eq.~(\ref{Kelvinlog}). Using direct numerical solution of the BdG equations it was found that the kelvon dispersion relation 
\begin{equation}
\omega^{\rm BdG}_{k} = \omega_0 + \frac{\hbar k^2}{2M}\left[\log\left( \frac{1}{kr_c}\right)\right],
\label{Kelvinquantumfix}
\end{equation} 
is shifted by $\omega_0$, which is the frequency of the fundamental kelvon \cite{Simula2008a}. As such, in the long-wave length limit, $k\to 0$, the kelvon has a non-zero frequency. 

Considering a rectilinear axisymmetric vortex of length $L$ with integer winding number $w$, the BdG equations (\ref{bog}) have stationary quasiparticle solutions of the form
\begin{align}
u_k({\bf r})&= u_k(r)e^{i [n  2\pi z/L + (\ell+w)\theta]} \notag \\ 
v_k({\bf r})&= v_k(r)e^{i [n 2\pi z/L + (\ell-w)\theta]},
\label{bdgmodes}
\end{align} 
where the integers $n$ and $\ell$ are, respectively, the principal and orbital angular momentum quantum numbers of the excitation, and $z$ and $\theta$ are the coordinates of the cylindrical coordinate system. 
 
For axisymmetric vortices, the kelvons have angular momentum quantum number $\ell=-w$ and quantised excitation frequencies $\omega_k$ and elementary excitation energies $E_k=\hbar\omega_k$. Within the linear response approximation, the condensate wave function perturbed by a kelvon is 
\begin{equation}
\psi({\bf r} ,t) = \left\{\psi({\bf r} ) +\epsilon \left[u_k({\bf r})e^{-i \omega_k t} +v^*_k({\bf r})e^{i \omega_k t}\right]\right\}e^{-i \mu t/\hbar} ,
\label{perturb}
\end{equation}
where $\epsilon$ determines the strength of the perturbation (kelvon population). Such kelvon quasiparticles \cite{Pitaevskii1961a} correspond to the classical Kelvin wave motion \cite{Kelvin1880a}. Fetter provides a clear theoretical discussion on the physics of the kelvon in BECs \cite{Fetter2004a}. When the quantised vortex is perturbed by kelvons by sufficiently large amplitude the vortex core is displaced from and orbits on a circular path around its own equilibrium position with angular frequency $\omega_k$ \cite{Movie}. Evidence for the existence of kelvons with great wave numbers on quantised vortices have been obtained via direct imaging in superfluid helium \cite{Fonda2014a} and in atomic Bose--Einstein condensates \cite{Bretin2003a,Lamporesi2017a}. 
 
The kelvon and the vortex are inseparable in the sense that one cannot exist without the other. For each vortex nucleated in the condensate wavefunction $\psi({\bf r}, t)$, a new low-energy quasiparticle state emerges in the spectrum of elementary excitations \cite{Simula2013a}. For a one vortex system the kelvon quasiparticle is localised inside the vortex core \cite{Isoshima1997a,Dodd1997a,Virtanen2001a}. Due to this localisation of the kelvon density within the vortex core, its excitation energy is very sensitive to the vortex core structure. 

Importantly, the $n=0$ kelvon quasiparticle also determines the experimentally observed orbital motion of a vortex in the centre of trapped Bose--Einstein condensates \cite{Anderson2000a,Bretin2003a,Hodby2003a,Freilich2010a,Serafini2015}. In typical cold atom experiments the kelvon, also called the anomalous mode or the vortex precession mode \cite{Isoshima1997a,Dodd1997a,Virtanen2001a,Simula2008a,Simula2008b,Feder2001a,Fetter2004a,Fetter2009a}, has the lowest excitation energy in the system $E_k <\hbar\omega_{\rm trap}$, where $\omega_{\rm trap}$ is the usual harmonic trapping frequency.

\subsection{Vortex spin}
For a singly quantised, charge $+\kappa$ vortex with $w=+1$, the equations (\ref{bdgmodes}) show that the fundamental $n=0$ kelvon with $\ell=-1$ has quasiparticle amplitudes $u_k({\bf r}) = u_k(r)$ and $v_k({\bf r})= v_k(r)e^{-i 2\theta}$. The Bogoliubov quasiparticle components at the vortex core, $r=0$, may thus be expressed as a 2-spinor
\begin{equation}
\left(\begin{matrix}
    u_k   \\  v_k   \\
\end{matrix} \right)
=
\left(\begin{matrix}
1  \\ e^{-i 2\theta} \\
\end{matrix} \right)
\equiv
\left(\begin{matrix}
1  \\ 0 \\
\end{matrix} \right),
\label{spinor3}
\end{equation}
which may be referred to as a spin-up kelvon. The BdG equations are invariant under the simultaneous symmetry operations $E_k \to -E_k^*$ and  
\begin{equation}
\left(\begin{matrix}
    u_k   \\  v_k   \\
\end{matrix} \right)
\to
\mathcal{K} \sigma_x
\left(\begin{matrix}
    u_k   \\  v_k   \\
\end{matrix} \right)
\label{spinor2}
\end{equation}
where $\mathcal{K}$ denotes complex conjugation and $\sigma_x=\left(\begin{matrix}
    0 \;\; 1   \\  1 \;\;0    \\
\end{matrix} \right)$ is a Pauli spin matrix. Application of this transformation to the kelvon yields 
\begin{equation}
\left(\begin{matrix}
    u_k   \\  v_k   \\
\end{matrix} \right)
=
\left(\begin{matrix}
e^{i 2\theta}  \\ 1 \\
\end{matrix} \right)
\equiv
\left(\begin{matrix}
0  \\ 1 \\
\end{matrix} \right)
\label{spinor}
\end{equation}
which may be referred to as a spin-down kelvon. 

\subsection{Vortex velocity dipole moment}

Measured with respect to the vortex-centric reference frame of the condensate wave function $\psi({\bf r}) = \psi(r)e^{i \theta}$ the perturbing particle and hole-like quasiparticle components $u_k({\bf r})$ and $v^*_k({\bf r})$  have phase windings $-1$ and $+1$, respectively, as is readily found by multiplying Eq.~(\ref{perturb}) by $e^{-i \theta}$. Dynamics causes the positions of these phase singularities to be spatially separated and leads to an intrinsic vortex velocity dipole moment (vVDM) of the vortex \cite{Klein2014a,Groszek2017a} due to the resulting kelvon induced dipolar superflow within the vortex core. The vVDM may be understood from the energetic perspective as the doubly charged phase singularity, $e^{\pm i 2\theta}$, in one of the quasiparticle components has a tendency to split into two spatially separated singly charged singularities via a critical point explosion \cite{Freund1999a}. Presumably, for the spin-up kelvons the vVDM is polarised in the direction of motion of the vortex where as for the spin-down kelvons the vVDM is antiparallel with respect to the velocity vector of the vortex. In addition to the charge, spin and vVDM, the kelvon quasiparticle is also the source of the inertial mass of the vortex.
 
\section{Vortex mass}

To determine the vortex mass, we begin with the Newton's second law
\begin{equation}
{\bf F}_v=M_v {\bf a}, 
\end{equation}
where ${\bf a}$ is the acceleration of the vortex and ${\bf F}_v$ is the force responsible for the acceleration. We take the constant of proportionality, $M_v$, as the definition of the inertial mass of the vortex.

To determine the inertial mass of the vortex we first consider a Bose--Einstein condensate trapped within a cylindrical hard-wall bucket potential of radius $R_\circ$ such that the condensate density $n({\bf r}, t)$ is practically constant everywhere in the fluid except within the vortex core and near the walls of the bucket. In this case, and at low temperatures, ${\bf F}_{\rm buoy} \approx 0$ and the vortex motion as observed in the laboratory reference frame is determined by the condition ${\boldsymbol v}^{\rm lab}_v\approx {\boldsymbol v}^{\rm lab}_s$. The full equation of motion in the laboratory frame of reference is
\begin{equation}
{\bf F}^{\rm lab}_{v} =  {\bf F}^{\rm lab}_{s}  + {\bf F}^{\rm lab}_{\rm buoy} = M_v{\bf a}^{\rm lab}.
 \label{forcebalancelab}
\end{equation}
The result in the absence of dissipation is that the vortex travels along a circle with a constant speed, constant acceleration and constant orbital angular frequency $\omega_v(r_v)$, the value of which depends on the radial position of the vortex. The acceleration vector in the laboratory frame then always points toward the centre of the circle. Therefore, if  ${\bf F}^{\rm lab}_{v} $ points radially outward direction, $M_v$ is negative and if ${\bf F}^{\rm lab}_{v} $ points in the same direction, toward the origin, as the acceleration vector, then $M_v$ is positive. 

In the uniform case, the circular motion of the vortex is due to the self-induced asymmetry of the superflow in the vicinity of the vortex. From the field theory perspective, this motion of the vortex may be viewed as the system's response to the broken continuous rotation symmetry. In this sense, the fundamental kelvon is a pseudo Nambu--Goldstone boson that aims to restore the broken rotation symmetry due to the presence of the vortex. Equivalently, the embedding flow ${\boldsymbol v}^{\rm lab}_s$ that drives the motion of the vortex can be viewed to be generated by an image vortex of opposite sign of circulation placed outside of the bucket. Directly integrating the Gross-Pitaevskii equation confirms that the observed frequency $\omega_v(r_v)\approx\omega_v(0) / [1-(r_v/R_\circ)^2]$ is consistent with the solution of the corresponding point-vortex model \cite{Groszek2017a}, which can be mapped onto the exactly soluble two-dimensional electrostatic problem of a charge inside a conducting ring. Once $M_v$ has been determined, it could be included in simulations of vortex dynamics that use point-vortex and vortex filament models, to include effects due to vortex inertia. Indeed, the first or the second equivalence in (\ref{forcebalancelab}) could be used to find the position of the vortex as a function of time. The former involves integrating the vortex velocity ${\boldsymbol v}_v(t)$ once, the latter requires integrating the acceleration ${\bf a}_v(t)$ twice to obtain ${\bf r}_v(t)$.

In order to relate the driving force and acceleration to the mass of the superfluid vortex, we make a transformation to a reference frame that rotates at the angular frequency $\omega_v(r_v)$ of the vortex. In such a frame, the vortex is stationary, ${\boldsymbol v}^{\rm rot}_v=0$. 
The equation of motion in the rotating frame of reference is
\begin{equation}
{\bf F}^{\rm rot}_{s}  + {\bf F}^{\rm rot}_{\rm buoy} = M_v{\bf a}^{\rm rot}
 \label{forcebalancerot1}
\end{equation}
which may be equivalently expressed as 
\begin{equation}
{\bf F}^{\rm lab}_{s}  + {\bf F}^{\rm lab}_{\rm buoy} ={\bf F}^{\rm lab}_{v} =  M_v{\boldsymbol \omega}^{\rm lab}_v\times{\boldsymbol v}^{\rm lab}_v,
 \label{forcebalancerot2}
\end{equation}
where the last term is due to centrifugal acceleration in the rotating frame.  The buoyancy force is the same in both frames since the frame transformation does not affect the observed condensate density in the vicinity of the vortex core and the embedding velocity field is the same in both frames of reference because it is produced by the image vortex, which maintains its position with respect to the actual vortex in all frames of reference. 

Substituting to (\ref{forcebalancerot2}) the expression for the vortex force in the laboratory reference frame
\begin{equation}
{\bf F}^{\rm lab}_{v}/L = -M \tilde{n}{\boldsymbol \kappa}\times {\boldsymbol v}^{\rm lab}_v
\end{equation}
we obtain the result for the velocity dependent vortex mass per unit length
$
M_v = \gamma (v_{\rm rel}) M_0= -2\pi \hbar\tilde{n}(r_v)L/\omega_v( r_v),
$
where $\gamma(v_{\rm rel}) $ is a factor that depends on the relative speed $v_{\rm rel}=|v_v-v_s|$ of the vortex with respect to the embedding flow.
The inertial rest mass per unit length $m_0=M_0/L$, with $\gamma(v_{\rm rel}) =1$ of the vortex is obtained by considering the limit of $r_v\to 0,v_v\to0$ and the result is
\begin{equation}
m_0=  \frac{2\pi\hbar n_0}{ \omega_k}
\label{restmass}
\end{equation}
and may also be expressed as  $m_0=M\rho_0 \kappa^2/ 2\pi E_k.$ Here $\omega_k =-\omega_v( r_v=0)$ is the fundamental kelvon frequency (the anomalous mode), whose excitation frequency in harmonically trapped BECs is known to be negative \cite{Feder2001a,Fetter2004a}.

In Eq.~(\ref{restmass}) $n_0=\tilde{n}(0)$ is the background condensate density at the origin in the absence of the vortex, $\rho_0=Mn_0$, and $\omega_k$ is the frequency of the kelvon quasiparticle, which equals the (negative of) orbital angular frequency of the vortex in the limit when the vortex approaches the centre of the bucket. Although $v_v\to0$ as $r_v\to 0$, the orbital angular frequency $-\omega_v(0)$ saturates to a non-zero value that equals the kelvon frequency $\omega_k$. Equation (\ref{restmass}) shows that the inertial mass of the vortex is indeed well defined and is determined microscopically by the excitation energy $E_k$ of the kelvon quasiparticle of the vortex. The kelvon frequency $\omega_k$ is determined self-consistently by the Bogoliubov--de Gennes equations, Eq.~(\ref{bog}), and it carries complete information of the influences on the vortex dynamics such as the near field and the internal structure of the vortex core, the far field away from the core, as well as the effects of quantum fluctuations and thermal atoms.

\section{Origin of the vortex mass}

Equation (\ref{restmass}) sheds light on the apparent issue of the logarithmic divergence of the incompressible kinetic energy associated with a \emph{static} vortex. In a Bose--Einstein condensate a standard calculation of the field mass yields $M_{\rm field}/L= E_{\rm field}/Lc_s^2\approx\frac{\pi\hbar^2}{g}\ln(R/r_c)$, where $R$ is the radius of the condensate and $r_c$ is approximately the size of the vortex core \cite{Pitaevskii1961a}. This suggests that the vortex mass would become infinite in the $r_c/R\to 0$ limit. We may also express Eq.~(\ref{restmass}) in the form $m_0= \frac{\pi\hbar^2}{g} \frac{2\mu}{\hbar\omega_k}$. For a harmonically trapped quasi-two-dimensional condensate, in the non-interacting $g\to0$ limit the chemical potential $\mu=\hbar\omega_{\rm trap}=|\hbar\omega_k|$ \cite{Fetter2009a} and $\ln (R/r_c)=\mathcal{O}(1)$ showing that our result and the logarithmically divergent mass definitions converge toward a finite value in this limit. However, in typical cold atom experiments $\mu/\hbar\omega_k\gg\ln(R/r_c)$. This means that a vortex in an interacting system is actually \emph{heavier} than the usual logarithmically divergent prediction. 

The reason is that the mass of the vortex $m_0$ is due to the mass $M_k$, defined below in Eq.~(\ref{kelvonmass}), of the kelvon quasiparticles. The situation is analogous to the case of an electron whose mass $M_e = eB/\omega_c$, where $e$ is the electric charge and $B$ is a constant magnetic field strength, is readily obtained from the measurement of its cyclotron frequency $\omega_c$ that relates the magnetic Lorentz force to the centrifugal force, \emph{cf.} Eq.~(\ref{restmass}). However, the energy $W=\frac{\epsilon_0}{2}\int |{\bf E}|^2d{\bf r}^3$ of the classical electromagnetic Coulomb field described by Maxwell's equations of classical electrodynamics, analogous to the kinetic energy $W=\frac{\rho}{2}\int |{\bf v}|^2d{\bf r}^3$ of $\psi({\bf r},t)$ in Eq.~(\ref{GP}), is divergent due to the singularity at the location of the electron and leads to the well known issues in the evaluation of the mass of the electron.

Equations (\ref{bog}), however, are the semi-classical analog 
\begin{equation}
\left(
\begin{matrix}
  -E_k+  \mathcal{H}    &  \Phi   \\
    \Phi^*   &  E_k + \mathcal{H}   \\
\end{matrix}
\right)
\left(
\begin{matrix}
    u_k   \\
    v_k   \\
\end{matrix}
\right)
=
\left(
\begin{matrix}
0  \\
0 \\
\end{matrix}
\right)
\label{dirac}
\end{equation}
of the Dirac equation and uncover the mass of the vortex. Indeed, Popov showed the equivalence of a two-dimensional system of phonons and vortices to relativistic electrodynamics \cite{Popov1973a}. The vortex core forms a harmonic oscillator potential in the condensate density, which in the vicinity of the vortex core, $r<r_c$, is 
\begin{equation}
n(r)\approx  n_0   \frac{r^2}{r^2+ 2r_c^2}, 
\end{equation}
where $r$ is the distance from the axis of the vortex. The vanishing of the condensate density at the vortex core allows the following substitutions to be made to the BdG equations (\ref{bog});
\begin{equation}
\Phi=\mathcal{V}M/M_k, 
\label{phiprox}
\end{equation}
\begin{equation}
E_k=(i\hbar \partial_t+\mu)M/M_k 
\label{haxprox}
\end{equation}
and
\begin{align}
\mathcal{H}=(\mathcal{M}({\bf r}, t) + \mu )M/M_k  ,
\label{maxprox}
\end{align}
where the semi-classical kelvon energy is
\begin{equation}
\mathcal{H}= -\frac{\hbar^2}{2M_k }\nabla^2+ V(r), 
\label{maxprox2}
\end{equation}
with the effective harmonic oscillator potential in the region $r\ll r_c$ defined by
\begin{equation}
V(r)=2gn(r) \frac{M}{M_k} =   gn_0   \frac{r^2}{r_c^2} \frac{M}{M_k} \equiv \frac{1}{2}M_k\omega^2_kr^2.
\label{kelvontrap}
\end{equation}
Equation~(\ref{kelvontrap}) may be solved for the mass of the kelvon with an approximation $E_k=M_k\omega_k^2r^2_c/2$, and results in 
\begin{equation}
M_k =\frac{\mu}{\hbar \omega_k}M,
\label{kelvonmass}
\end{equation} 
consistent with the approximations made in Eq.~(\ref{phiprox}), Eq.~(\ref{haxprox}) and  Eq.~(\ref{maxprox}). Replacing the relativistic mass-energy $\mathcal{H}=Mc^2$ in the Dirac equation by the semi-classical energy $p^2/2M+V$ of Eq.~(\ref{maxprox2}) shows the connection between the Dirac and the Dirac--Bogoliubov--deGennes equation (\ref{dirac}).

The kelvon mass naturally agrees with an estimate based on the Heisenberg uncertainty relation $\Delta x_k\Delta p_k\approx \hbar/2$, with uncertainties in the vortex position $\Delta x_k\approx r_c$ and momentum $\Delta p_k\approx M_k \omega_k r_c$, which yields $M_k \approx \hbar/2\omega_kr^2_c=M \mu/E_k $. At zero temperature the number of kelvons $N_k$ in a vortex of length $L$ in the vortex core is $N_k=M_0/M_k =L/4a$, where $a$ is the $s$-wave scattering length.
Thus we find that the mass and the total energy per unit length of the vortex are large mostly due to the \emph{zero point motion of the vortex}. Although Eq.~(\ref{restmass}) seems to admit an infinite vortex mass in the form of a zero-energy kelvon mode, vanishing kelvon frequency is not consistent with Eq.~(\ref{Kelvinquantumfix}) and would be unphysical in light of the Heisenberg uncertainty principle \cite{Byckling1965a}. Furthermore, a vortex moving on a cylinder of any size can never be static \cite{Guenther2017a}, which shows the importance of boundary conditions even in systems that extend to infinity.

\subsection{Bose and Fermi superfluids}

\begin{figure}[!t]
\centering
\includegraphics[width=0.8\columnwidth]{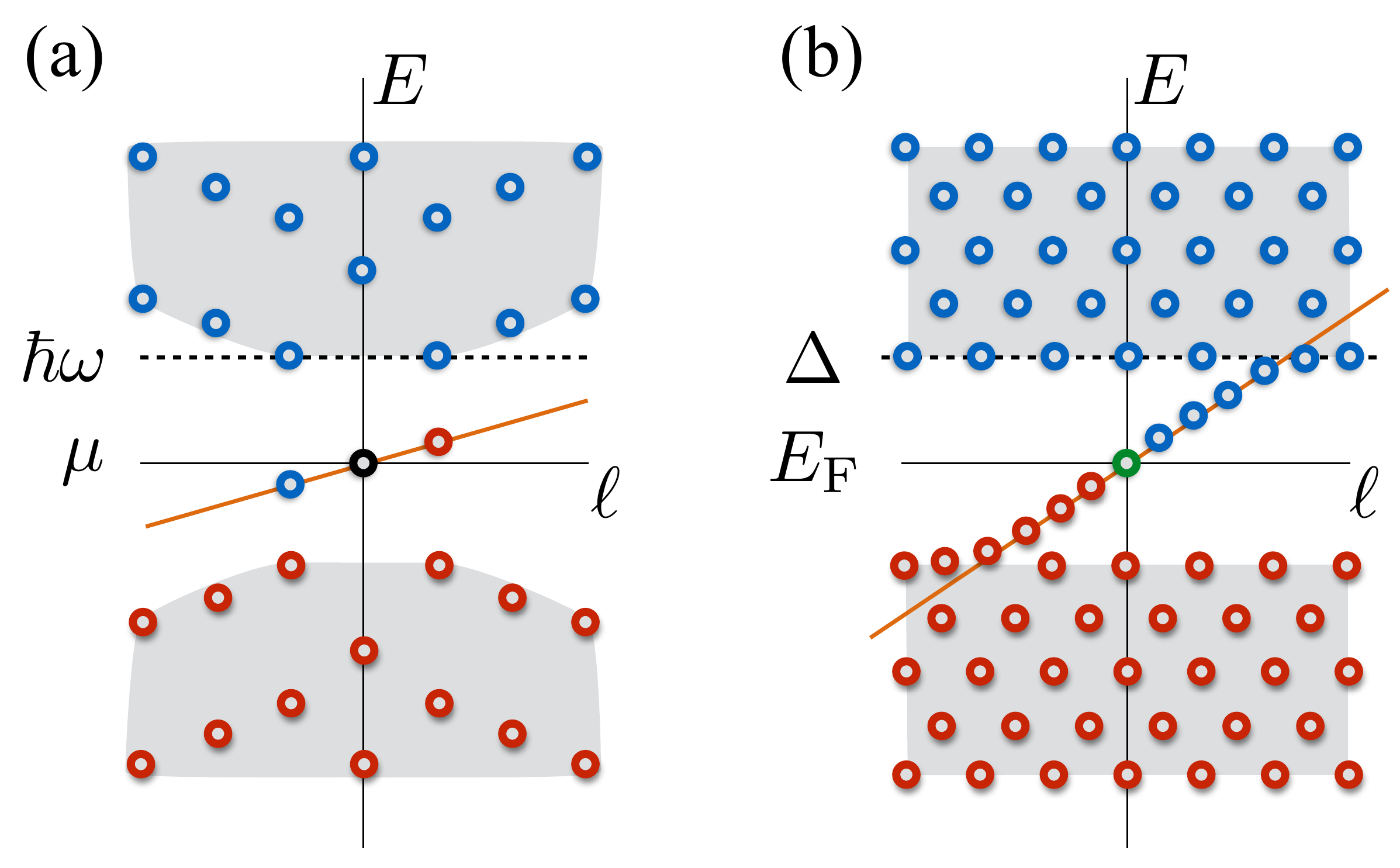}
\caption{
(color online) Illustration of the elementary BdG excitation energies $E$ as functions of orbital angular momentum quantum number $\ell$ for a vortex state in a scalar Bose superfluid (a) and in a topological Fermi superfluid (b). The slope of the orange line in (a) and (b) is $\omega_k$ and $\omega_{\rm F}$, respectively. The zero mode in (a) is the Goldstone boson (black circle) and in (b) it is the Majorana fermion (green circle). The red circles denote the hole-like quasiparticles states and the blue circles denote the particle-like quasiparticle states. The quasiparticle modes within the gap in (a) and (b) due to the presence of the vortex are the kelvons and the CdGM modes, respectively. 
}\label{kuva}
\end{figure}

Figure \ref{kuva} (a) illustrates the elementary BdG quasiparticle excitation spectrum for a vortex state in a simple Bose--Einstein condensate. The excitation energies shown as a function of angular momentum quantum number $\ell$ are measured with respect to the chemical potential $\mu$. The zero energy mode (black circle) with $\ell=0$ is the Goldstone boson corresponding to the macroscopically occupied condensate mode. The blue and red circles denote the particle and hole-like quasiparticle eigenmodes, respectively. In the Bose system the canonical commutation relations result in the hole-like modes (red circles) to have negative norm and therefore they are discarded from the self-consistent sums in Eqns~(\ref{rho}) and (\ref{delta}).

For harmonically trapped bosons, the excitation energy gap equals the harmonic oscillator level spacing $\hbar\omega$ and the lowest lying excitation mode is the kelvon with $\ell=-1$ and $|E_k|\ll \hbar\omega$. In two-dimensional systems, there is only one kelvon mode in the spectrum and its energy $E_k\approx -0.1 \hbar\omega$ in typical harmonically trapped systems \cite{Freilich2010a} is negative with respect to the chemical potential $\mu$. The slope of the orange line equals the magnitude of the kelvon frequency $\omega_k$. In harmonically trapped BECs this kelvon frequency within the Thomas--Fermi approximation is \cite{Svidzinsky2000a}
\begin{equation}
\omega_k^{\rm TF} = -\frac{3\hbar}{2MR_\perp^2} \log\left( \frac{R_\perp}{r_c}\right), 
\end{equation}
where $R_\perp$ is the radial Thomas--Fermi radius of the condensate. Substituting this estimate for the kelvon frequency to Eq.~(\ref{restmass}) yields a vortex mass estimate
\begin{equation}
M_{\rm TF} = -\frac{4\pi}{3} \frac{n_0 R_\perp^2L}{\log\left( R_\perp/r_c\right)} M.
\label{MTF}
\end{equation}

Baym and Chandler \cite{Baym1983a} considered vortices in a helium II---a strongly interacting bosonic superfluid---and obtained a vortex mass
\begin{equation}
M_{\rm BC} = \pi \rho_s r_c^2 L,
\end{equation}
which is equivalent to the bare mass $M_{\rm bare}$. The Baym--Chandler mass may also be expressed in terms of the inertial mode frequency $\omega_I=2\Omega$ of a rapidly rotating vortex lattice rotating at orbital angular frequency $\Omega$ as $M_{\rm BC} = 2\pi\hbar \rho_sL/M\omega_I $ \cite{Baym1983a}. Comparing this with our result, Eq.~(\ref{restmass}), shows that the essential difference between them is that the Baym--Chandler mass involves the inertial mode frequency $\omega_I\gg \omega_k$ of a vortex lattice instead of the kelvon frequency $\omega_k$ of a single vortex, and therefore results in a much smaller value for the vortex mass than our kelvon based result. In rapidly rotating vortex lattices, the inertial mode frequency, corresponding to the standard inertial mode of a rotating fluid, is orders of magnitude greater than the Tkachenko mode frequency \cite{Baym1983a,Baym2003a,Simula2004a} although the individual vortices in both of these modes execute Kelvin wave like motion. This suggests that for vortex lattices the total mass of the vortex matter might be obtained by replacing the Kelvin wave frequency in Eq.~(\ref{restmass}) by the Tkachenko mode frequency.

For cold atom systems with chemical potential $\mu=gn_0\gg\hbar\omega_k$, Eq.~(\ref{restmass}) yields $m_0=4M_{\rm bare} \mu/L\hbar\omega_k$, such that the inertial vortex mass is in general much greater than the bare mass. In uniform two-dimensional traps $\omega_v(0)= \hbar / MR_\perp^2 $ such that Eq.~(\ref{restmass}) yields $M_{\rm uni}\approx -2NM$, where $N$ is the number of atoms in the condensate. The magnitude of the inertial vortex mass in this case equals twice the total gravitational mass of the condensate.

Figure \ref{kuva} (b) illustrates the elementary BdG quasiparticle excitation spectrum for a vortex state in a topological Fermi superfluid measured with respect to the Fermi energy $E_{\rm F}$. The zero energy mode (green circle) is a Majorana fermion quasiparticle corresponding to one of the Caroli--deGennes--Matricon (CdGM) modes localised within the vortex core. In BCS superfluids the CdGM modes may be viewed as Andreev bound states trapped by the normal state vortex core. For a topological Fermi superfluid the CdGM modes are 
\begin{equation}
\omega_{\rm CdGM} = -\left(\ell-\frac{w_{+1} +1}{2}\right )\omega_{\rm F} + \left(n-\frac{w_{+1}+1}{2}\right )\omega_1,
\end{equation}
where $\omega_{+1}$ is the integer winding number of the dominant component of the chiral pair potential \cite{Mizushima2008a}.
The level spacing of the CdGM modes, the slope of the orange line in (b), is $\omega_{\rm F} \approx \Delta^2/E_{\rm F}\ll \omega_1\approx \Delta$. In Fermi systems, the gap may contain more than one core modes even in two-dimensional systems, in contrast to the Bose case. 

The Zwierlein group studied the vortex mass in Fermi systems \cite{Yefsah2013a,Ku2014a} obtaining a generic formula 
\begin{equation}
M_{\rm BF} = -\frac{4\pi}{2\gamma+1} \frac{n_0 R_\perp^2L}{\log\left( R_\perp/r_c\right)} M,
\end{equation}
where the polytropic index $\gamma=1$ for BECs and $\gamma=3/2$ for a Fermi gas in the BCS limit and at unitarity. For $\gamma = 1$ this result is identical to Eq.~(\ref{MTF}).

The Kopnin mass \cite{Kopnin1978a,Sonin2013a} 
\begin{equation}
m_{\rm Kop} =\frac{\pi\hbar n_0}{\omega_{\rm F}}
\end{equation} 
in fermonic superfluids is expressed in terms of $\omega_{\rm F}$ and is equivalent, up to the factor of 2, to the kelvon based vortex mass in Bose systems since in both cases the cyclotron frequency in the denominator equals the slope of the vortex core localised quasiparticle modes. Note that the absent factor of 2 in the Kopnin mass could be recovered by considering that in single quantum vortices of BCS superconductors the angular momentum states are shifted by 1/2 such that the CdGM level spacing in such a case actually corresponds to 2 times the lowest CdGM frequency. These considerations show that the inertial mass of a vortex in Bose and Fermi superfluids has the same origin---the quasiparticle modes localised within the vortex core.

\subsection{Geometric phase of the vortex}
In a uniform superfluid the geometric Berry phase 
\begin{equation}
\gamma(C) = i \oint_C \langle \Psi |\nabla_{\bf R} \Psi \rangle \cdot d{\bf R}
\label{Berry}
\end{equation}
and the vortex force are related by \cite{Thouless1996a}  
\begin{equation}
\gamma(C)= 2\pi w N(C) = \int_C \frac{({\bf f}_{\rm v}\times {\boldsymbol v}_v)\cdot d{\bf a}}{\hbar v^2_v},
\label{Berryman}
\end{equation}
where $C$ denotes the closed orbital path of the vortex, $w$ is the winding number and $N(C)$ is the number of atoms enclosed by $C$. In this case the velocity dependent vortex mass and Berry phase are related by
\begin{equation}
\frac{\gamma(C)}{M_v(C)}\hbar L  = \int_C {\boldsymbol \omega}_v(C)\cdot d{\bf a}.
\label{Berrymass}
\end{equation}
This profound connection between the topological charge of the quantised vortex and its dynamical behaviour also reveals that on exchanging the positions of two $w=1$ vortices the system acquires a phase of $2\pi$ per atom enclosed by the exchange path. In two-dimensional superfluids with more complex order parameter structure such as spinor BECs, the vortices are in general anyons and may possess non-Abelian exchange phases.  

\section{Experimental prospects}
Superfluid Bose and Fermi gases provide a promising experimental platform for precision measurements of the Berry phase, Magnus force, inertial mass and dipole moment of a quantised vortex. An off-centre vortex in such systems can be created and its orbital angular frequency $\omega_v(r_v)$ as a function of its position observed. The measured orbital frequencies in both Bose \cite{Anderson2000a,Bretin2003a,Hodby2003a,Freilich2010a,Serafini2015} and Fermi \cite{Yefsah2013a,Ku2014a} gases have been found to be in good agreement with the mean-field theory \cite{Fetter2009a,Scott2011a,Liao2011a,Koens2013a,Efimkin2015a,Toikka2017a}. Extrapolating such frequency data to $r_v=0$ yields the kelvon frequency $\omega_k$. This, in combination with a measurement of the condensate density, yields the inertial rest mass $m_0$ of the vortex. A systematic study of the vortex mass could be performed using cold atom gases trapped in bucket potentials by varying the radial position and the core size of the vortex in the superfluid. Uniform trapping geometries suitable for such experiments are already being used by several groups \cite{Eckel2016a,Gauthier2016a,Mukherjee2016a,Tempone2017a}. Controlling the vortex size and position could be achieved by first nucleating several vortices in the system using standard methods and then waiting until only one vortex remains. Further wait time causes the radial position of the remaining vortex to increase due to dissipative effects and allows for dialling the radial position of the vortex. Alternatively, external optical pinning potentials could potentially be used for moving the vortex in the desired radial position in the trap. The particle density of the condensate determines the healing length and thereby the size of the vortex core and can be adjusted by varying the number of atoms that form the condensate. Tuning the effective inter-species interaction in a two-species condensate where one species hosts the vortex and the other species plugs the vortex core could also be used to intrinsically modify the vortex core size and its kelvon frequency, and thereby the mass of the vortex.  

Fortuitously, the same experimental setup for measuring the kelvon frequencies could also be used to measure the Berry phase of a vortex in a superfluid for the first time. The Berry phase of a vortex could be obtained using its discretised form 
\begin{equation}
\gamma(C) = - {\rm Im}\left(\log\prod_i^P \int \psi^*({\bf r}, t_i) \psi({\bf r}, t_{i+1})dv\right),
\label{discreteBerry}
\end{equation}
where the closed path $C$ of the vortex is sampled at $P$ different vortex positions and the integration is over all space. Experimentally, the integrands in Eq.~(\ref{discreteBerry}) could potentially be extracted from an interferogram of two wave functions acquired at time intervals $\delta t=t_{i+1} -t_1$ during which the vortex is allowed to travel a distance of the order of its core diameter. The dynamical phase that accumulates at the rate determined by the chemical potential must also be carefully accounted for. The measured value of $\gamma(C)$ should be compared with the predicted value  $2\pi w N(C)$, where the number of enclosed atoms $N(C)$ could perhaps be measured to shot-noise limited accuracy \cite{Gajdacz2016a}. Using the measured values of the Berry phase and the vortex speed, the vortex force $f_v$ can be obtained using the Eq.~(\ref{Berryman}), and the Magnus force $f_{\rm Mag}$ can be obtained directly by measuring the condensate density that allows $f_{\rm buoy}$ to be calculated. Using Eq.~(\ref{forcebalancemagnus}) $f_s$ can be measured. The vVDM could potentially be detected using velocity selective Bragg scattering techniques. 

\section{Classical limit}

According to Bohr's correspondence principle, quantum systems in the limit of large quantum numbers are expected to behave classically. For vortices, this amounts to replacing in the equation (\ref{restmass}) the quantum of circulation $\kappa=h/M$ by $\Gamma = w \kappa$ and the condensate mass density $M n_0$ by the fluid density $\rho$ to yield 
\begin{equation}
M_{\rm vortex} =   \frac{\Gamma  \rho}{ \omega_{\rm K}}L .
\label{classmass}
\end{equation}
In macroscopic systems such as bath tub vortices the effective winding number $w$ would be comparable to the Avogadro number. In 1956 Hall and Vinen modelled a quantised vortex as a rotating cylinder filled with the fluid it is immersed in \cite{Hall1956a}. They showed that the motion of such a cylinder with mass $M_{\rm cyl}$ that displaces a fluid of mass $M_{\rm flu}$, involves a constant velocity perpendicular to an applied force and an oscillatory motion at the frequency 
\begin{equation}
\omega_{\rm cyl}= \rho \Gamma L / (M_{\rm cyl}+M_{\rm flu}). 
\label{cylinder}
\end{equation}
Identification of the frequency $\omega_{\rm cyl}$ of such motion with the fundamental frequency of a Kelvin wave of a vortex and the total mass $M_{\rm cyl}+M_{\rm flu}$ with the inertial mass of the vortex recovers the result of Eq.~(\ref{classmass}).

\begin{table}[!t]
\caption{\label{table}  Typical inertial masses of vortices.}
\begin{ruledtabular}
\begin{tabular}{lcccc}
   & $\Gamma$ (m$^2$/s)& $\rho $ (kg/m$^3$)& $\omega_{\rm K}$ (rad/s)& $M_v/L$ (kg/m)\\ 
  \hline\\
        superconductor & $4\times10^{-4}$  & $2\times 10^{-6}$   & $1\times10^{12}$ & $5\times10^{-22}$  \\
            quantum gas & $5\times10^{-9}$  & $1\times10^{-5}$    & $3\times10^{1}$ & $3\times10^{-15}$    \\   
            neutron star & $4\times10^{-7}$  & $1\times 10^{17}$ & $2\times10^{20}$ & $2\times10^{-10}$   \\  
             superfluid $^{4}$He & $1\times10^{-7}$  & $1\times10^{2}$    & $6\times10^{0}$ & $2\times10^{-6}$ \\                         
                        water & $6\times10^{-3}$  & $1\times10^{3}$    & $1\times10^{-1}$ & $6\times10^{1}$    \\
                         air & $8\times10^{3}$  &  $1\times10^{0}$    & $1\times10^{0}$ & $1\times10^{4}$               
\end{tabular}
\end{ruledtabular}
\end{table}

We will next consider a classical vortex ring instead of a columnar vortex to avoid the issue of boundary conditions at the ends of the vortex. Consider thus a vortex ring produced by a piston with a circular orifice of radius $R$ as in the experiments by Sullivan \emph{et al.} \cite{Sullivan2008a}. By invoking momentum conservation, the product of the inertial mass of the vortex ring and its velocity must be equal to the momentum of the fluid set in motion by the piston. This yields
\begin{equation}
M_v V_R = \rho \Gamma \pi R^2. 
\label{mom balance}
\end{equation}
Therefore, the inertial mass of a vortex ring of length $L=2\pi R$ is
\begin{equation}
M_v =   \frac{\Gamma  \rho}{ \omega^\circ_{\rm K}}L ,
\label{classmassO}
\end{equation}
where 
\begin{equation}
\omega^\circ_{\rm K} = \frac{2V_R }{R}
\end{equation}
is the magnitude of the angular frequency of the fundamental, $n=0$, Kelvin mode on a vortex ring. This result agrees with the quantum mechanical expression Eq.~(\ref{restmass}). However, there is a discrepancy regarding the constant $\eta$. In Eq.~(\ref{classmassO}) $\eta=2$, where as Eq.~(\ref{KelvinThom}) predicts it to be either 0 or 1/2. One possible explanation could be that the curvature of the vortex ring may require a corrective factor due to the contribution of Kelvin waves of higher wave numbers. Moreover, the Eq.~(\ref{KelvinThom}) is derived for hollow vortex rings where as the cores of the vortex rings in the experiment were filled with fluid.  Note also that in the experiment \cite{Sullivan2008a}, the ratio of induced mass to the mass of the fluid trapped by the vortex ring was $0.65$, whereas our definition for the inertial vortex mass is all-inclusive and contains the mass of the moving fluid. It thus seems that a detailed experimental study of the Kelvin wave frequencies as a function of the velocity of a vortex ring is warranted and could potentially be achieved using the method of Kleckner and Irvine \cite{Kleckner2013a}.

Using Eq.~(\ref{classmass}) it is straight forward to estimate typical values of the inertial mass of a vortex for various physical systems shown in Table~I. The circulation $\Gamma$ for classical systems may be estimated as $\Gamma = 2\pi v r$, where $v$ is the fluid velocity at distance $r$ from the centre of the vortex, where as for quantum systems $\Gamma =\kappa= h/M$ is determined by the Planck's constant and the mass of the particle forming the superfluid such as a Cooper pair of electrons or neutrons or an atom. 

In Table I, we have used the following estimates: For a superconductor we choose, $M=2M_e$, $\rho = 2 M_e \times 10^{24} /{\rm m}^3$, $\omega_{ k} = 10({\rm km/s}) / 50 {\rm nm} \approx \Delta^2/E_{\rm F}$ \cite{Embon2017a}. For an atomic BEC we choose, $M=M$($^{87}$Rb), $\rho = 10^{-5} \; {\rm kg}/{\rm m}^3$, $\omega_{ k} = 2\pi \times 4$Hz \cite{Freilich2010a}. For superfluid helium II we choose, $M=M$($^{4}$He), $\rho = 125\; {\rm kg}/{\rm m}^3$, $\omega_{k} = 2\pi \times 1$Hz \cite{Fonda2014a}. For a neutron star we choose, $M=$ amu, $\rho = 10^{17}\; {\rm kg}/{\rm m}^3$, $\omega_{k} = 1 {\rm MeV}/\hbar$ \cite{Elgaroy2001a}. For air we consider a `Fujita 1' tornado and choose, $\Gamma = 8\times 10^{3} \; {\rm m^2}/{\rm s} $, $\rho = 10^{1}\; {\rm kg}/{\rm m}^3$, and $\omega_{\rm K} = 1$ rad/s. For water we consider a bath tub vortex and choose, $\Gamma = 6\times10^{-3} \; {\rm m^2}/{\rm s} $, $\rho = 10^{3}\; {\rm kg}/{\rm m}^3$, $\omega_{\rm K} = 0.1$ rad/s.

\section{Conclusions}
In conclusion, we have studied the inertial vortex mass in a superfluid. We find that the total inertial mass of a quantised vortex in a superfluid Bose--Einstein condensate, Eq.~(\ref{restmass}), is determined microscopically by the condensate density and the self-consistent elementary excitation frequency of the kelvon quasiparticle. The result is all-inclusive in the sense that, in contrast to previous mass estimates for bosonic superfluids, it includes all contributions to the inertial mass of a vortex such as the effects of the near and far-field superflows and the core filling substances such as non-condensate atoms. The vortex is heavy and, protected by the Heisenberg uncertainty principle, its mass does not suffer from logarithmic divergencies. In addition to the inertial mass of the quantised vortex, the kelvon quasiparticle is also responsible for the charge, spin and vortex velocity dipole moment (vVDM) of the quantised vortex. We find that the inertial mass of a vortex in Bose and Fermi superfluids has the same origin---quasiparticles localised within the vortex core. Considering the classical limit of large quantum numbers we obtain a relationship between the Kelvin waves and the inertial mass of classical vortices and vortex rings.

The inertial mass of a vortex may have relevance to many areas of physics including two-dimensional quantum turbulence, gravitational wave emission from neutron stars, topological quantum computing and high-temperature superconductivity. Avalanches of anomalously heavy vortices in rotating neutron stars that glitch could perhaps result in a continuous gravitational wave signal detectable by future gravitational wave detectors \cite{Melatos2015a}. In high-energy states of two-dimensional quantum turbulence the vortex particles may undergo condensation transition \cite{Valani2018a}---a phenomenon which may be influenced by the mass of the vortices. The ability to control the inertial mass of a vortex by varying the frequency of the kelvons has potential for applications. In Bose--Einstein condensates such controlling of the inertial mass of quantised vortices could be achieved using vortex pinning laser beams that shift the excitation energies of the kelvons \cite{Simula2008a}. In high temperature superconductors it would be desirable to be able to reduce the motion of the vortices since vortices of greater inertial mass would require greater de-pinning force enabling the material to withstand stronger supercurrents and potentially higher critical temperatures \cite{Embon2017a}. In a topological quantum computer based on braiding of non-Abelian vortex anyons the goal is the opposite. The faster the vortices can be moved around, the faster the gate operations necessary for the quantum information processing can be achieved. Thus, depending on the specific application, it may be desirable to make the vortices as heavy or as light as possible.

\begin{acknowledgments}
I am grateful to Victor Galitski, Andrew Groszek, Kris Helmerson, David Paganin and Martin Zwierlein for useful discussions. This work was financially supported by the Australian Research Council via Discovery Projects No.  DP130102321 and No. DP170104180.
\end{acknowledgments}

\bibliographystyle{apsrev}

\end{document}